\documentclass[a4paper,11pt]{article}
\pdfoutput=1
\usepackage{jheppub}
\usepackage{subcaption}
\usepackage{todonotes}
\newcommand{\ve}{\varepsilon}
\newcommand{\om}{\omega}
\newcommand{\Za}{Z\alpha}

\newcommand{\BS}{{e\gamma}}
\newcommand{\PP}{{e\bar e}}
\newcommand{\ep}{\epsilon}

\newcommand{\Li}{\mathrm{Li}}

\renewcommand{\Re }{\operatorname{Re}}
\newcommand{\G}{{F}}

\title{Charge asymmetry in the spectra of bremsstrahlung and pair production}

\author{Petr A. Krachkov}
\author{and Roman N. Lee}

\affiliation{Budker Institute of Nuclear Physics, Novosibirsk 630090, Russia}

\emailAdd{r.n.lee@inp.nsk.su}
\emailAdd{p.a.krachkov@inp.nsk.su}

\abstract{We calculate the first Coulomb correction to the spectra of two processes: the electron bremsstrahlung and electron-positron photoproduction in the Coulomb field.
We show that, in contrast to the results obtained in the Born approximation and in the high-energy limit, the obtained corrections for these two process are not related by the crossing symmetry substitutions.
The found corrections determine the leading contribution to the charge asymmetry in these processes. We use modern multiloop methods based on the IBP reduction and on the differential equations for master integrals. The results are presented in terms of classical polylogarithms. We provide both the threshold and the high-energy asymptotics of the obtained expressions and compare them with available results.}

\allowdisplaybreaks
\begin{document}

\maketitle
\flushbottom

\section{Introduction}
The electron bremsstrahlung and photoproduction of electron-positron pairs in the Coulomb field of the nucleus are fundamental QED processes. First papers devoted to this subject appeared already in 1930s. In particular, in 1934 Bethe and Heitler have derived the spectra of both processes in the Born approximation \cite{BetheHeitler1934}. The total energy loss in bremsstrahlung and the total cross section of pair production in the Born approximation have been calculated in the same year by Racah \cite{Racah1934a,Racah1934}.

Exact, in both energy and nucleus charge, pair production spectrum has been investigated much later in Refs. \cite{OMoOl1968,OMoOl1973}. The result had a very complicated form involving multiple sums and Appell double hypergeometric function $F_2$. Using this form the authors numerically tabulated the spectrum and the total cross section for energies $\lesssim 10$MeV. There is no analogous result for the spectrum of bremsstrahlung. Apart from Refs. \cite{OMoOl1968,OMoOl1973}, the Coulomb corrections, i.e., the higher-order terms in the parameter $\Za$ (here $Z$ is the nucleus charge number, $\alpha\approx 1/137$ is the fine structure constant) have been investigated mostly in the low- and  high-energy limit.
The low-energy asymptotics of the bremsstrahlung spectrum has been considered by Sommerfeld \cite{sommerfeld1931beugung}.
The earliest results for the threshold asymptotics of the pair production spectrum were obtained by Nishina, Tomonaga and Sakata in  Ref. \cite{NTS1934} using the Furry-Sommerfeld-Maue approximation.
However it appears that this approximation is not sufficient and the correct results were obtained only recently \cite{Krachkov:2022fgt}.
The leading and subleading high-energy asymptotics of the differential cross sections and spectra of both processes have been obtained exactly in the parameter $\Za$  in Refs. \cite{BethMax1954,DavBeMa1954} and \cite{Lee2004,Lee2005}, respectively.

For light nuclei the parameter $\Za$ is small and one can use the perturbation theory with respect to this parameter. The leading Born approximation is invariant under the replacement of electron with positron and vice versa. For the bremsstrahlung it means that electron and positron radiate exactly the same in this approximation. For the pair production it means that the electron spectrum is symmetric under the change $\ve_-\leftrightarrow \ve_+$ (here $\ve_{\pm}$ are the electron/positron energies). The first Coulomb correction which we consider in the present paper determines the charge asymmetry in these two processes. For the bremsstrahlung process this correction determines the difference between the radiation from electron and that from positron. For the pair photoproduction cross section the first Coulomb  correction determines the leading anti-symmetric contribution to the energy spectrum of the produced pair.

Recently, the multiloop methods have been used in Ref. \cite{Lee2021} to calculate the first Coulomb correction to the energy loss in bremsstrahlung. In the present paper we use the multiloop methods to calculate the first Coulomb correction to the energy spectra of bremsstrahlung and pair production.

\section{Cross sections}

For the bremsstrahlung process we define $\ve,\ \ve^\prime$ and $\om^\prime=\ve-\ve^\prime$ to be the energies of the initial electron, final electron and final photon, respectively. Similarly, for the pair production process we define $\ve_-,\ \ve_+$ and $\om=\ve_-+\ve_+$ to be the energies of the  final electron, final positron and initial photon. The physical regions are determined by the inequalities
\begin{align}
	\text{bremsstrahlung:}&\quad m<\ve^\prime<\ve\,,\label{eq:reg_BS}\\
	\text{pair production:}&\quad \ve_\pm>m\,.\label{eq:reg_PP}
\end{align}
The two processes are related to each other by the crossing symmetry
\begin{equation}\label{eq:crossing}
	\ve^\prime \leftrightarrow \ve_-,\quad \ve\leftrightarrow-\ve_+,\quad \om^\prime\leftrightarrow-\om\,.
\end{equation}
Note that these substitution rules are too ambiguous to determine the analytic continuation of the differential cross section from one physical region to another. Moreover, such a continuation necessarily exists only for the amplitudes of the processes. It is known that the spectra of bremsstrahlung and pair production processes in the Born approximation as well as in the high-energy approximation are also related by the substitutions \eqref{eq:crossing}. Moreover, the consideration of Ref. \cite{OlseMax1959} can be erroneously taken as a proof of exact symmetry between the two processes. However we shall see below that already first corrections to the spectra considered here have completely different forms and can not be obtained by not only a naive substitution \eqref{eq:crossing}, but also by any carefully chosen analytic continuation.

The spectrum of photons in the bremsstrahlung process reads
\begin{equation}
	\frac{d\sigma^{\BS}}{d\om^\prime}
	=\frac{1}{2|\boldsymbol{p}|}\int\overline{\textstyle\sum}\left|M\right|^2\delta\left(\om^\prime-|\boldsymbol{k}'|\right)d\Phi\,,\qquad
	d\Phi=2\pi\delta\left(\ve-\ve_{\boldsymbol{p}^\prime}-|\boldsymbol{k}'|\right)\frac{d\boldsymbol{k}'}{(2\pi)^32|\boldsymbol{k}'|}\frac{d\boldsymbol{p}^\prime}{(2\pi)^32\ve_{\boldsymbol{p}^\prime}}\,.
\end{equation}
Here we use the notation $\ve_{\boldsymbol{p}^\prime}=\sqrt{\boldsymbol{p}^{\prime2}-m^2}$.

The perturbative expansion of the matrix element $M$ starts with the term $\propto e\Za$. As we are interested in the first Coulomb  correction, we write
\begin{align}
	M&=e\Za \left[M_0 + \Za M_1+\ldots\right],\\
	|M|^2&=4\pi\alpha(\Za)^2\left\{\overline{\textstyle\sum}|M_0|^2+2\Za\overline{\textstyle\sum}\Re M_1M_0^*+\ldots\right\}.
\end{align}
where dots denote omitted higher-order terms in the parameters $\alpha$ and/or $\Za$. Consequently we have
\begin{equation}
	\frac{d\sigma^{\BS}}{d\om^\prime}=\frac{d\sigma_B^{\BS}}{d\om^\prime}+\frac{d\sigma_C^{\BS}}{d\om^\prime}+\ldots\,,
\end{equation}
where
\begin{equation}
	\frac{d\sigma_B^{\BS}}{d\om^\prime}=\frac{\alpha(\Za)^2}{(2\pi)^4|\boldsymbol{p}|}\int d\mu_{\BS}\overline{\textstyle\sum}\left|M_0^\BS\right|^2
\end{equation}
is the Born cross section and
\begin{equation}\label{eq:sigma_C}
	\frac{d\sigma_C^\BS}{d\om^\prime}=\frac{2\alpha(\Za)^3}{(2\pi)^4|\boldsymbol{p}|}\int d\mu_\BS\overline{\textstyle\sum}\Re\left[ M_1^\BS \left(M_0^{\BS}\right)^*\right]
\end{equation}
is the first Coulomb correction that we consider in the present paper. Here
\begin{equation}
	d\mu_\BS=\delta\left(\om^\prime-|\boldsymbol{k}'|\right)\delta\left(\ve^\prime-\ve_{\boldsymbol{p}^\prime}\right)\frac{d\boldsymbol{k}'}{2|\boldsymbol{k}'|}\frac{d\boldsymbol{p}^\prime}{2\ve_{\boldsymbol{p}^\prime}}
\end{equation}
is the integration measure.

\begin{figure}
	\centering
	\begin{subfigure}{0.5\textwidth}
		\includegraphics[width=\textwidth]{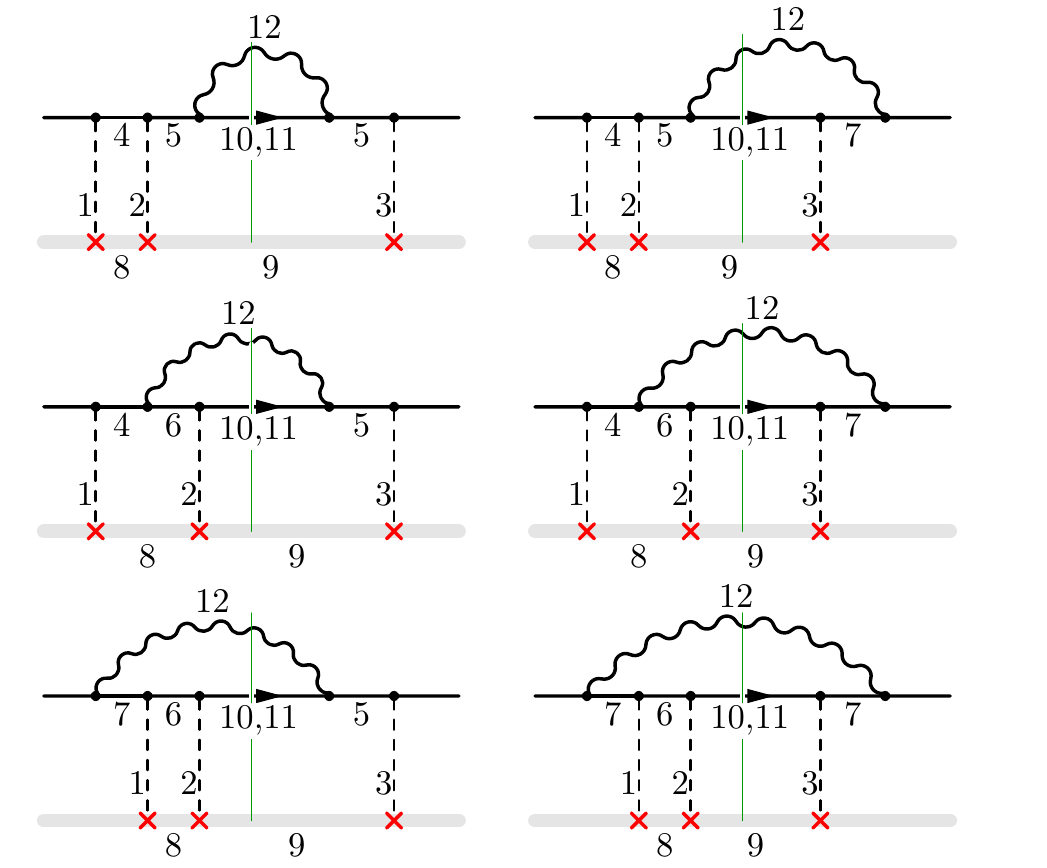}
		\caption{Cut diagrams that determine $\frac{d\sigma_C^\BS}{d\om^\prime}$.}
		\label{fig:DiagramsBS}
	\end{subfigure}\hfill
	\begin{subfigure}{0.5\textwidth}
		\includegraphics[width=\textwidth]{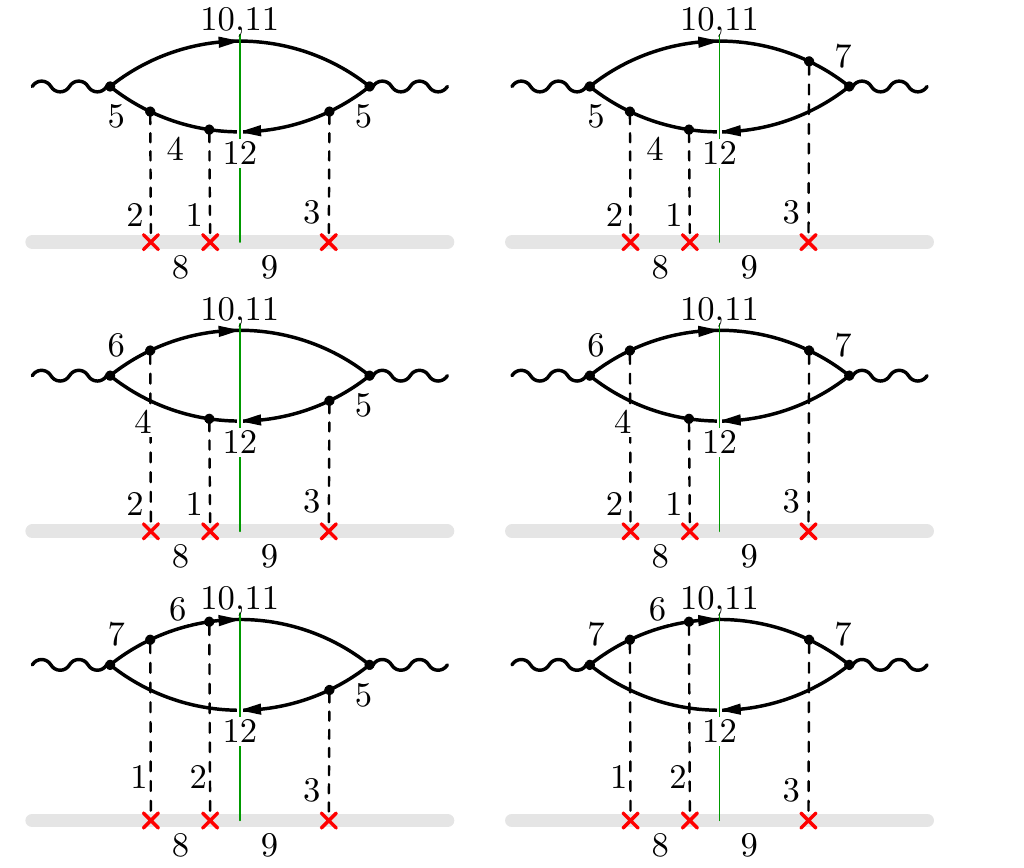}
		\caption{Cut diagrams that determine  $\frac{d\sigma_C^\PP}{d\om^\prime}$.}
		\label{fig:DiagramsPP}
	\end{subfigure}
\caption{Cut diagrams for first Coulomb correction in bremsstrahlung and pair production spectra. Numbers on the diagrams correspond to the enumeration of functions $D_k$, Eq. \eqref{eq:Ds_BS}, (left) and $\widetilde{D}_k$, Eq. \eqref{eq:Ds_PP}, (right).}
\label{fig:DiagramsBSPP}
\end{figure}

Similar formulas hold for the pair production process:
\begin{align}
	\frac{d\sigma^\PP}{d\ve_+}&=\frac{d\sigma_B^\PP}{d\ve_+}+\frac{d\sigma_C^\PP}{d\ve_+}+\ldots\,,\\
	\frac{d\sigma_B^{\PP}}{d\ve_+}&=\frac{\alpha(\Za)^2}{(2\pi)^4\om}\int d\mu_{\PP}\overline{\textstyle\sum}\big|M_0^\PP\big|^2\,,\\
	\frac{d\sigma_C^\PP}{d\om^\prime}&=\frac{2\alpha(\Za)^3}{(2\pi)^4\om}\int d\mu_\PP\overline{\textstyle\sum}\Re\left[ M_1^\PP \left(M_0^{\PP}\right)^*\right]\,,\\
	d\mu_\PP&=\delta\left(\ve_+-\ve_{\boldsymbol{p}_+}\right)\delta\left(\ve_--\ve_{\boldsymbol{p}_-}\right)\frac{d\boldsymbol{p}_+}{2\ve_{\boldsymbol{p}_+}}\frac{d\boldsymbol{p}_-}{2\ve_{\boldsymbol{p}_-}}
\end{align}
Graphically, $\frac{d\sigma_C^\BS}{d\om^\prime}$ and $\frac{d\sigma_C^\PP}{d\ve_+}$ can be represented as a sum of diagrams depicted in Fig. \ref{fig:DiagramsBS} and in Fig. \ref{fig:DiagramsPP}, respectively. Note that we have ordered the diagrams in these figures in a way consistent with the crossing symmetry \eqref{eq:crossing}. Namely, each diagram in  Fig. \ref{fig:DiagramsPP} can be obtained from the corresponding diagram in Fig. \ref{fig:DiagramsBS} by opening wide the cut photon line and closing the external fermion legs into a cut positron line.

\section{Calculation of master integrals}\label{sec:IBP_DE}
We use the dimensional regularization with $d=4-2\ep$ and consider the integral family
\begin{equation}
	\label{eq:j_BS}
	j(n_{1},\ldots,n_{12})=\Re\int\frac{dqdk^\prime dp^\prime }{\pi^{3d/2}}
	\prod_{k=1}^{7}D_{k}^{-n_{k}}\times
	\prod_{k=8}^{12}\frac{\delta^{(n_k-1)}\left(-D_{k}\right)}{(n_k-1)!}\,,
\end{equation}
where
\begin{gather}
	D_1  =q^{2}\,,\quad D_2 =\left(Q-q\right){}^{2}\,,\quad D_3 =Q^{2}\,,\quad D_4 =\left(p-q\right)^{2}-m^2+i0\,, \nonumber\\
	D_5 =\left(k^\prime+p^\prime \right)^{2}-m^2\,, \quad D_6  =\left(p-k^\prime -q\right){}^{2}-m^2+i0\,,\quad
	D_{7}  =\left(p-k^\prime \right)^{2}-m^2\,,\nonumber\\
	D_8=q\cdot n\,, \quad D_9=Q\cdot n\,,\quad D_{10} =p^\prime \cdot n-\ve^\prime\,,\quad D_{11}=p^{\prime 2}-m^2\,,\quad D_{12}=k^{\prime 2}\,.
	\label{eq:Ds_BS}
\end{gather}
Here $Q=p-p^\prime -k^\prime$ and $n=(1,\boldsymbol 0)$ is the time ort. The cut denominators $D_{8-12}$ correspond to the on-shell condition for the emitted photon ($D_{12}$), the zero energy transfer to the heavy nucleus ($D_{8}$ and $D_{9}$), the on-shell condition for the final electron ($D_11$), and the $\delta$-function fixing the electron energy ($D_{10}$).
Note that we have indicated $+i0$ prescription only in those denominators in Eq. \eqref{eq:Ds_BS} which may otherwise turn to zero in the integration region\footnote{The remaining denominators in Eq. \eqref{eq:Ds_BS} have definite sign in the whole integration region: $D_5>0$ and $D_{1,2,3,7}<0$.}.The Coulomb correction $\frac{d\sigma_C^\BS}{d\om}$ in Eq. \eqref{eq:sigma_C} is expresses via integrals $j(n_1,\ldots n_{12})$ for which $n_{8-12}=1$ and at least one of $n_4,\ n_5,\ n_6,\ n_7$ is non-positive.

Similarly, for the pair production we consider the family
\begin{equation}
	\label{eq:j_PP}
	\tilde{j}(n_{1},\ldots,n_{12})=\Re\int\frac{dqdp_+dp_-}{\pi^{3d/2}}
	\prod_{k=1}^{7}\widetilde{D}_{k}^{-n_{k}}\times
	\prod_{k=8}^{12}\frac{\delta^{(n_k-1)}\left(-\widetilde{D}_{k}\right)}{(n_k-1)!}\,,
\end{equation}
where
\begin{gather}
	\widetilde{D}_1  =q^{2}\,,\quad \widetilde{D}_2 =(\widetilde{Q}-q){}^{2}\,,\quad \widetilde{D}_3 =\widetilde{Q}^{2}\,,\quad \widetilde{D}_4 =\left(p_++q\right)^{2}-m^2+i0\,, \nonumber\\
	\widetilde{D}_5 =\left(p_--k\right)^{2}-m^2\,, \quad \widetilde{D}_6  =\left(k-p_+-q\right){}^{2}-m^2+i0\,,\quad
	\widetilde{D}_7  =\left(k-p_+\right)^{2}-m^2\,,\nonumber\\
	\widetilde{D}_{8}=q\cdot n\,, \quad \widetilde{D}_{9}=\widetilde{Q}\cdot n\,,\quad  \widetilde{D}_{10} =p_- \cdot n-\ve_-\,,\quad \widetilde{D}_{11}=p_-^2-m^2\,,\quad \widetilde{D}_{12}=p_+^2-m^2\,.
	\label{eq:Ds_PP}
\end{gather}
Here $\widetilde{Q}=k-p_- -p_+$. Note that our choice of the functions $D_k$ and $\widetilde{D}_k$ is well adjusted to the crossing symmetry relation \eqref{eq:crossing}. Namely, all of them except the last ones ($D_{12}$ and $\widetilde{D}_{12}$) pass to each other, $D_k\leftrightarrow \widetilde{D}_k$, under \eqref{eq:crossing}.

Making the IBP reduction \cite{ChetTka1981,Tkachov1981} with \texttt{LiteRed} \cite{LiteRed2013,LiteRed2021}, we reveal 59 master integrals for each case (bremsstrahlung and pair production) and construct differential equations for them \cite{Kotikov1991,Remiddi1997}. We choose the master integrals for pair production in a symmetric fashion, so that the symmetry $\ve_+\leftrightarrow\ve_-$ is a one-to-one mapping of these integrals. The master integrals for the bremsstrahlung are chosen in a way consistent with crossing symmetry \eqref{eq:crossing}. Due to our judicious choice of denominators, \eqref{eq:Ds_BS} and \eqref{eq:Ds_PP}, this requirement simply means that for each pair production master integral $\widetilde{j}(n_1,\ldots,n_7,1,1,1,1,1)$ the integral ${j}(n_1,\ldots,n_7,1,1,1,1,1)$ is a bremsstrahlung master integral.

\begin{figure}
		\includegraphics[width=0.4\textwidth]{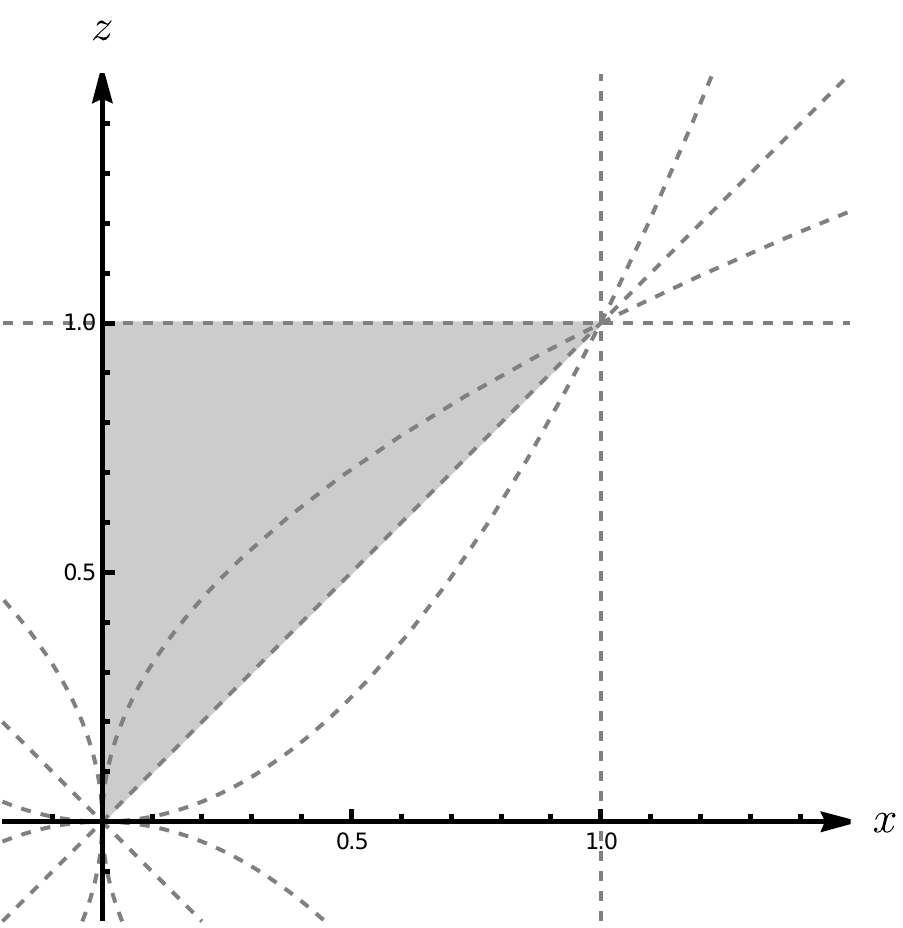}
\hfill
		\includegraphics[width=0.4\textwidth]{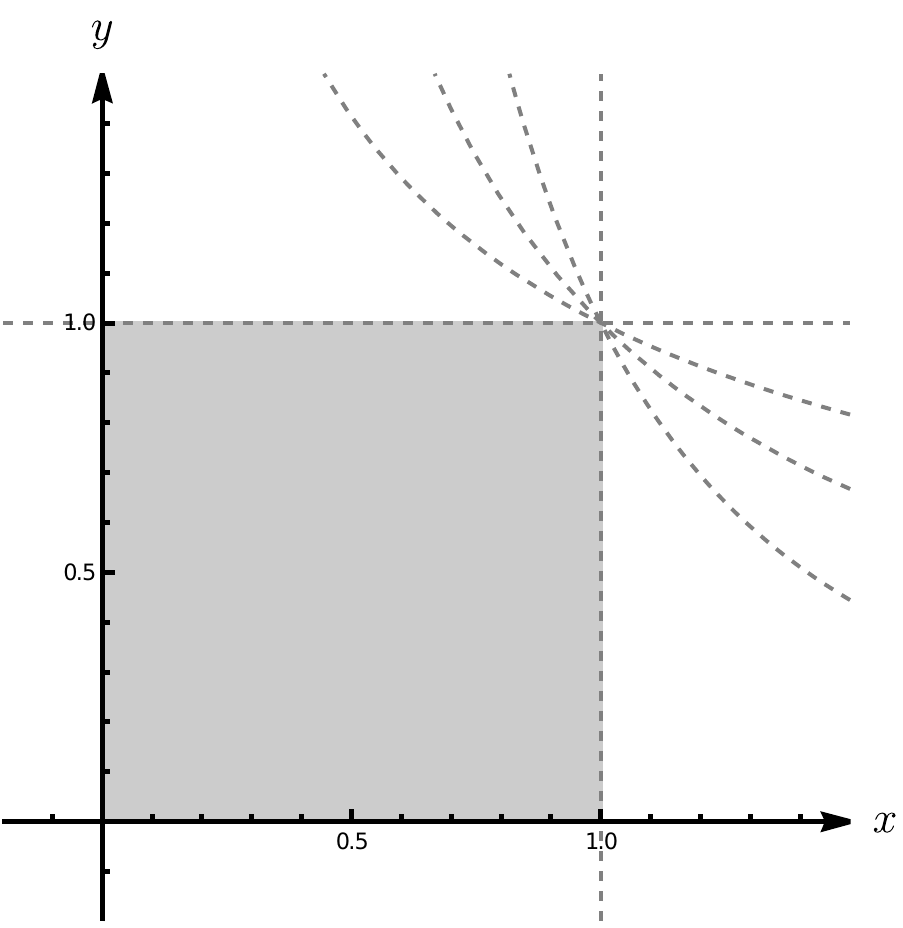}
	\caption{Physical region (grayed) and singularities (dashed curves) of the differential system for bremsstrahlung (left) and pair production (right) master integrals.}
	\label{fig:reg_BSPP}
\end{figure}

Using \texttt{Libra}\footnote{Note that \texttt{Libra} is able to treat the multivariate case.}, \cite{Libra2020}, we reduce the system to $\epsilon$-$d\!\log$-form \cite{Henn2013,Lee2014}. In order to do this, we introduce the following new kinematic variables:
\begin{align}
	\text{bremsstrahlung:}&\qquad x=\sqrt{\tfrac{\ve^\prime-m}{\ve^\prime+m}}\,,\quad
	z=\sqrt{\tfrac{\ve-m}{\ve+m}}\,,\\
	\text{pair production:}&\quad\ x=\sqrt{\tfrac{\ve_--m}{\ve_-+m}}\,,\quad
	y=\sqrt{\tfrac{\ve_+-m}{\ve_++m}}\,.
\end{align}
In new variables the crossing symmetry corresponds to the exchange rules
$z\leftrightarrow -1/y\,,\ x\leftrightarrow x$.
The physical regions are
\begin{align}
	\text{bremsstrahlung:}&\quad 0<x<z<1\,,\label{eq:reg_bs}\\
	\text{pair production:}&\quad 0<x,y<1\,.\label{eq:reg_pp}
\end{align}
The resulting differential system for bremsstrahlung has the form
\begin{gather}\label{eq:de_bs}
	d \boldsymbol{J} =\ep dM\, \boldsymbol{J}\,,\quad dM=\sum_{k=1}^{12}M_k\, d\ln P_k\,,\\
	\label{eq:alphabet}
	(P_1,\ldots,P_{12})=(x,z,1-x,1-z,1+x,1+z,x-z,x+z,x^2-z,x^2+z,z^2-x,z^2+x)\,,
\end{gather}
where $M_k$ are some constant matrices. The plot of singular curves corresponding to the alphabet \eqref{eq:alphabet} is shown in the left plot of Fig. \ref{fig:reg_BSPP}. Note that the singularity $x-z^2=0$ traverses the physical region.

The case of pair production process is very similar.
 Reducing the differential system to $\ep$-form, we obtain a similar system \eqref{eq:de_bs} where the alphabet now has the form
\begin{equation}
(P_1,\ldots,P_{12})=(x,y,1-x,1-y,1+x,1+y,1-xy,1+xy,1-x^2y,x^2+z,z^2-x,z^2+x)\,,
\label{eq:alphabet1}
\end{equation}
The physical region and singular curves for pair production are shown in Fig. \ref{fig:reg_BSPP}, right.

\subsection{Boundary conditions}
For the bremsstrahlung case we fix the boundary conditions by considering the asymptotics of the master integrals in the region $x\ll z\ll 1$. More specifically, we first consider the limit $x\to 0$ and determine which coefficients of the asymptotic expansion in $x$ are to be calculated. Those coefficients depend on $z$ and we construct the differential system for them using the method described in Ref. \cite[Sec. 2]{Lee:2020obg}. Reducing this system to $\ep$-form, we determine which coefficients of the asymptotic expansion in $z$ should be fixed. We find these coefficients by a direct integration method. Note that the obtained boundary constants are such that the special solution (in contrast to general solution) has no branching at $x-z^2=0$ in the physical region, as it should be.

For the pair production case fixing the boundary conditions is done by passing from $x,\ y$ to $t,\ r$ via $x=tr,\ y=t/r$ and considering the asymptotics $t\to 0$ at fixed $r$. In contrast to the bremsstrahlung case here it appears to be easy to obtain the necessary asymptotic coefficients as functions of $r$ by direct integration methods, without messing with the differential equations in this variable.

\subsection{Crossing symmetry}

\begin{figure}
	\centering
	\includegraphics[width=0.7\linewidth]{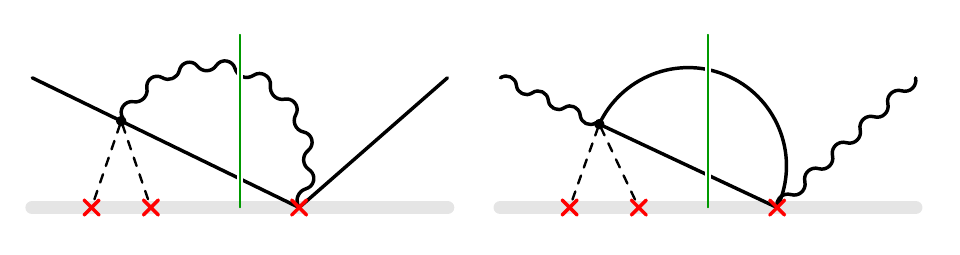}
	\caption{Example of a pair of master integrals, $j_{20}$ and $\widetilde{j}_{20}$, which are not related by analytical continuation.}
	\label{fig:example}
\end{figure}

A remarkable observation that we can make about our results is that most of the master integrals do not obey the crossing symmetry. Let us demonstrate this fact on the example of one master integral for bremsstrahlung and its counterpart for pair production. These master integrals, $j_{20}$ and $\widetilde{j}_{20}$ in our enumeration, are shown in Fig. \ref{fig:example}.
These integrals are finite in $d=4$.  We are going to demonstrate that already $\ep^0$ terms can not be related via analytical continuation. Thus, below in this section we put $d=4$.
The corresponding analytic expressions for the integrals are
\begin{align}
	I(\ve,\ve^\prime)&\stackrel{\text{def}}{=}4\pi\ve^\prime\om^\prime j_{20} = \frac1{\pi^3}\left\langle\int  \frac{d\boldsymbol{q}}{(\boldsymbol{p}-\boldsymbol{p}^\prime -\boldsymbol{k}^\prime-\boldsymbol{q})^2\boldsymbol{q}^2}\right\rangle_{\boldsymbol{p}^\prime,\, \boldsymbol{k}^\prime}\label{eq:j20_BS}
	\\
	\widetilde{I}(\ve_+,\ve_-)&\stackrel{\text{def}}{=}4\pi\ve_-\ve_+ \widetilde{j}_{20} = \frac1{\pi^3}\left\langle\int  \frac{d\boldsymbol{q}}{(\boldsymbol{k}-\boldsymbol{p}_- -\boldsymbol{p}_+-\boldsymbol{q})^2\boldsymbol{q}^2}\right\rangle_{\boldsymbol{p}_-,\, \boldsymbol{p}_+}\label{eq:j20_PP}
\end{align}
Here $\langle\ldots \rangle_{\boldsymbol{v}}$ denotes the averaging over the angles of the vector $\boldsymbol{v}$, i.e., the integration with the measure $\int \frac{d\Omega_{\boldsymbol{v}}}{4\pi}$.
Note that the right-hand sides of Eqs. \eqref{eq:j20_BS} and \eqref{eq:j20_PP} obviously do not change upon the averaging over the angles of the remaining vector, $\boldsymbol{p}$ for bremsstrahlung and $\boldsymbol{k}$ for the pair production. Comparing the integrands, we see that they are related by the crossing substitutions \eqref{eq:crossing}. This property persists after taking the integral over $\boldsymbol{q}$:
\begin{align}
I(\ve,\ve^\prime)&= \left\langle  \frac1{|\boldsymbol{p}-\boldsymbol{p}^\prime -\boldsymbol{k}^\prime|}\right\rangle_{\boldsymbol{p}^\prime,\, \boldsymbol{k}^\prime,\, \boldsymbol{p}}\label{eq:j20_BS_1}
\\
\widetilde{I}(\ve_+,\ve_-)&= \left\langle  \frac1{|\boldsymbol{k}-\boldsymbol{p}_- -\boldsymbol{p}_+|}\right\rangle_{\boldsymbol{p}_-,\, \boldsymbol{p}_+,\, \boldsymbol{k}}\label{eq:j20_PP_1}
\end{align}
At this point one might speculate that $\widetilde{I}(\ve_+,\ve_-)=I(-\ve_+,\ve_-)$, where the right-hand side is understood as a suitable analytic continuation of Eq. \eqref{eq:j20_BS_1} to the region of negative first argument. However, it is not the case. We use the following formula\footnote{Note that when $a,\,b,\,c$ satisfy triangle inequality, the result is different, $A(a,b,c)=(2ab+2ac+2bc-a^2-b^2-c^2)/(4abc)$.}
\begin{equation}\label{eq:A}
	A(a,b,c)\stackrel{\text{def}}{=}\left\langle  \frac1{|\boldsymbol{a}+\boldsymbol{b}+\boldsymbol{c}|}\right\rangle_{\boldsymbol{a},\,\boldsymbol{b},\,\boldsymbol{c}} =
	\frac1{a} \qquad \text{if}\quad a>b+c.
\end{equation}
It is easy to derive this formula by recalling that $\frac1{\sqrt{1+2t x +t^2}}$ is a generating function for the Legendre polynomials.
Using Eq. \eqref{eq:A} we obtain
\begin{equation}
	I(\ve,\ve^\prime) =\frac1{\sqrt{\ve^2-m^2}}\,,\quad
	\widetilde{I}(\ve_+,\ve_-) =\frac1{\ve_++\ve_-}\,.
\end{equation}
It is obvious that these two functions can not be related to each other by analytical continuation. In particular, $\widetilde{I}(\ve_+,\ve_-)$ is a single-valued meromorphic function.

We find similar incompatibility with the crossing symmetry as defined in Eq. \eqref{eq:crossing} in most of the master integrals as well as in the final result.

\section{Results}
Our final result for the bremsstrahlung energy-dependent cross section has the form
\begin{equation}
\frac{d\sigma^{\BS}}{d\om}=\frac{d\sigma_B^{\BS}}{d\om}+\frac{d\sigma_C^{\BS}}{d\om} +O(\alpha^2\eta^2,\alpha\eta^4)\,,
\end{equation}
where
\begin{multline}\label{eq:BS_born}
	\frac{d\sigma_B^{\BS}}{d\om'} =\frac{\alpha(\Za)^2}{\om'p^2}\Bigg\{
	\frac{4pp'}{3m^2}-2\frac{\ve \ve'}{m^2}\frac{p^2 + p'^2}{p p'}
	- 2\frac{\ve' p'\ln{r_z}}{p^2}- 2\frac{\ve p \ln{r_x}}{p'^2}- 4\ln{r_x}\ln{r_z}\\
	-2\ln{r_{x/z}}\left(\frac{8 \ve \ve'}{3 m^2}+
	\frac{\om'^2(\ve^2 \ve'^2+p^2 p'^2+m^2\ve \ve')}{m^2p^2 p'^2}
	-\frac{(\ve \ve'+p^2)\om'}{p^3}\ln{r_z}+\frac{(\ve \ve'+p'^2)\om'}{p'^3}\ln{r_x}\right)
	\Bigg\}
\end{multline}
is the Born cross section \cite{BetheHeitler1934}, and
\begin{multline}\label{eq:BS}
	\frac{d\sigma_C^{\BS}}{d\om'} =
	\frac{\pi\,\alpha(Z\alpha)^3}{p^2\om'}
	\Big\{
	c_1\, \left[\G(r_{x/z},r_x)-\G(r_{x/z},r_{x^2/z})+\ln{r_x}\ln{r_{x^2/z}}\right]
	+c_2 \G(r_x,r_{x/z})\\
	+c_3\, \left[\G(r_{x},r_{x^2/z})-\G(r_{x},r_{z})+\ln{r_{x/z}}\ln{(r_z/r_{x^2/z})}\right]
	+c_4\,\left[\Li_2(x)-\Li_2(-x)\right]
	-c_5\Re\G(r_{x},-1)\\
	+c_6\ln{r_x}\ln{r_{x/z}}
	+c_7\ln{r_{z}}\ln{r_{x/z}}
	-c_8\ln{r_x}\ln(r_x r_z^2)
	+c_9\ln{r_{x/z}}
	-c_{10}\ln{r_z}
	+c_{11} \ln{r_x}
	+c_{12}
	\Big\}\,,
\end{multline}
is the result of the present calculation. Here $r_x=
\tfrac{1-x}{1+x}
$ and the function $F$ is defined as
\begin{align}
	\G(r,\tilde{r}) &= \tfrac12\Li_2(1-r\tilde{r})+\tfrac12\Li_2(1-\tfrac r{\tilde{r}})
	-\tfrac12\Li_2(1-\tfrac{\tilde{r}}r)-\tfrac12\Li_2(1-\tfrac1{\tilde{r}r})\nonumber\\
	&=
	\Li_2(1-r\tilde{r})+\Li_2(1-\tfrac{r}{\tilde{r}})+\tfrac12\ln^2{r}+\tfrac12\ln^2{\tilde{r}}
	\,.
\end{align}
Note that from this definition it follows that $\G(r,\tilde{r})=\G(r,1/\tilde{r})=-\G(1/r,\tilde{r})$. The coefficients $c_k$ are rational functions of $z,x$ or, alternatively, of $\ve,\,p,\,\ve',\,p'$. Their explicit form is presented in appendix \ref{sec:appendix}.

For the pair production cross section we have
\begin{equation}\label{eq:PP_born}
	\frac{d\sigma^{\PP}}{d\om}=\frac{d\sigma_B^{\PP}}{d\om}+\frac{d\sigma_C^{\PP}}{d\om} +O(\alpha^2\eta^2,\alpha\eta^4)\,,
\end{equation}
where
\begin{equation}\label{eq:PP}
	\frac{d\sigma_B^{\PP}}{d\ve_-} =
	\left(\frac{p^2d\sigma_B^{\BS}}{\om'^2d\om'}\right)_{ (\ve',p',\ve,p,\om',z)\to(-\ve_+,p_+,\ve_-,p_-,-\om, -1/y)}\,,
\end{equation}
and
\begin{multline}
	\frac{d\sigma_C^{\PP}}{d\ve_-}=
	\frac{\pi\,\alpha(Z\alpha)^3}{\om^3}
	\bigg\{
	\tilde{c}_1\, \left[\G\left(r_y, r_{x\,y^2}\right)-\G\left(r_y,r_{x\,y}\right)-\ln {r_{x\,y}} \ln {r_{x\,y^2}}-2 \ln {r_x} \ln {r_y r_{x\,y}}\right] \\
	+\tilde{c}_2\, \left[\G\left(r_{x\,y},1\right)-\G\left(r_{x\,y},r_{x^2\,y}\right)+\ln {r_x}\ln{r_{x^2\,y}}\right]
	+\tilde{c}_3\,\left[\G\left(r_{x\,y},1\right)-2 \G\left(r_{x\,y},r_x\right)+\ln{r_y}\ln{r_x} \right]\\ +\tilde{c}_4\,\left[\Li_2(y\,x)-\Li_2(-y\,x)\right]
	-\tilde{c}_5 \ln {r_y} \ln {r_{x\,y}}
	+\tilde{c}_6\,\ln ^2{r_{x\,y}}
	+\tilde{c}_7 \ln {r_{x\,y}}
	+\tilde{c}_8 \ln {r_y}
	+\tilde{c}_9
	\bigg\}
	\,,\\
	-\left\{(\ve_-,p_-,x)\leftrightarrow (\ve_+,p_+,y)\right\}
\end{multline}
The explicit form of the coefficients $\tilde{c}_k=\tilde{c}_k(\ve_-,p_-,\ve_+,p_+)$ is presented in appendix \ref{sec:appendix}.

We attach the \textit{Mathematica} files \textsf{BSresult} and \textsf{PPresult} which contain Eqs. \eqref{eq:BS} and \eqref{eq:PP} in computer-readable form. The content of these files can be used as follows
\begin{itemize}
	\item The code\\
	\texttt{\small\{eq43, eq44, eqA1, ep2xz\} = Get["BSresult"];\\
		eq43//.eq44/.eqA1/.ep2xz/.\{x->0.3,z->0.4\} (*==>  133.445*al$^4$Z$^3$/m$^3$*)
	}\\gives the value of $\frac{d\sigma_C^{\BS}}{d\om'}$ at $x=0.3,\ z=0.4$.
	\item The code\\
	\texttt{\small \{eq47, eq44, eqA2, ep2xy\} = Get["PPresult"];\\
		(\#-(\#/.\{x->y,y->x\}))\&[eq47//.eq44/.eqA2/.ep2xy]/.\{x->0.3,y->0.9\}\\(*==>  0.146448*al$^4$ Z$^3$/m$^3$*)
	}\\gives the value of $\frac{d\sigma_C^{\PP}}{d\ve_-}$ at $x=0.3,\ y=0.9$.
\end{itemize}

\subsection{Asymptotics}
Let us present asymptotics of the obtained corrections to the spectra. We start from high-energy asymptotics. The first three terms of the high-energy asymptotics of the bremsstrahlung spectrum read
\begin{multline}\label{eq:asy_BS}
	\left(\frac{\pi\alpha(Z \alpha)^3}{\ve^2\om}\right)^{-1}\frac{d\sigma_C^{\BS}}{d\om} =
	\frac{\pi^2}{4m}\left(\tfrac{2\om'^2}{\ve\,\ve'}+3\right)(\ve+\ve')
	-\bigg\{
	\tfrac{\pi^2(\ve^2-\ve'^2)^2}{2\ve^2\ve'^2}+\left[\tfrac{\ve'}{\ve} \left(\ln{\tfrac{2\ve}{m}}-1\right)+\tfrac{\ve}{\ve'}\ln{\tfrac{2\ve'}{m}}\right]^2
	\\
	-\tfrac{8(\ve^2-\ve'^2)}{3\ve\,\ve'}\ln{\tfrac{2\ve\,\ve'}{\om' m}}
	+\tfrac{3\ve^3-3\ve^2\ve'+3\ve\,\ve'^2+\ve'^3}{\ve\,\ve'^2}\left(2\ln{\tfrac{2\ve'}{m}}-1\right)
	+2\tfrac{\ve^2-\ve'^2}{\ve^2}\ln{\tfrac{\om'}{\ve'}}\ln{\tfrac{\ve}{\ve'}}
	+\tfrac{19\ve^2+3\ve\,\ve'+11\ve'^2}{3\ve\,\ve'}
	\\
	-2\tfrac{\ve^2}{\ve'^2}\Re\text{Li}_2\left(1+\tfrac{\om'}{\ve}\right)
	+2\tfrac{2\ve^2-\ve'^2}{\ve^2}\text{Li}_2\left(\tfrac{\ve}{\ve'}\right)
	\bigg\}
	+\frac{\pi ^2 m\,\left(\ve^2 -\ve'^2\right) \left(\ve^3-\ve'^3 \right)}{2 \ve ^3 \ve'^3}
	+\mathcal{O}(m^2)
\end{multline}

For the pair production spectrum we have
\begin{multline}\label{eq:asy_PP}
	\left(\frac{\pi\alpha(Z \alpha)^3}{\om^3}\right)^{-1}\frac{d\sigma_C^{\PP}}{d\ve_-}=
	\frac{\pi^2 }{4 m}\left(\tfrac{2\om^2}{\ve_+\ve_-}-3\right)(\ve_+-\ve_-)
	+\bigg\{
	\tfrac{\pi ^2 \ve_+^2}{2 \ve_-^2}-\tfrac{2 \ve_+ }{\ve_-}\ln {\tfrac{2 \ve_-}{m}}-\tfrac{\ve_-^2+\ve_+^2}{\ve_-^2}\ln ^2{\tfrac{2 \ve_-}{m}}\\
	-\tfrac{ \ve_+\left(9 \om+\ve_- \right) }{3 \ve_-^2}\left(2\ln {\tfrac{2 \ve_- \ve_+}{\om\, m}}-1\right)+\tfrac{2 \ve_-^2 }{\ve_+^2}\Re\text{Li}_2\left(1+\tfrac{\om}{\ve_-}\right)+2 \text{Li}_2\left(-\tfrac{\ve_+}{\ve_-}\right)
	-\left(\ve_-\leftrightarrow \ve_+\right)\bigg\}\\
	+\frac{\pi ^2 m\,\left(\ve_+^2-\ve_-^2\right) \left(\ve_-^3+\ve_+^3\right)}{2 \ve_-^3 \ve_+^3}
	+\mathcal{O}(m^2)
	\,,
\end{multline}

The leading terms in Eqs. \eqref{eq:asy_BS} and \eqref{eq:asy_PP} are in agreement with the results of Refs. \cite{Lee2005} and \cite{Lee2004}, respectively, where the $\mathcal{O}\left(m/\ve\right)$ correction to the spectra were calculated exactly in the parameter $\Za$.

The low-energy asymptotics for bremsstrahlung and pair production processes, $v\sim v'\ll1$ and $v_+\sim v_-\ll 1$, respectively, reads
\begin{align}\label{eq:asy_BS_low}
	\left(\frac{\pi\alpha(Z \alpha)^3}{m^3}\right)^{-1}\frac{d\sigma_C^{\BS}}{d\omega}&=
	\frac{64\operatorname{arctanh} (v'/v)}{3 v^3 v' (v+v')}+\mathcal{O}(v^{-3})
	\,,\\
	\label{eq:asy_PP_low}
	\left(\frac{\pi\alpha(Z \alpha)^3}{m^3}\right)^{-1}\frac{d\sigma_C^{\PP}}{d\ve_-}&=
	\tfrac{1}{6}(v_+-v_-)(v_+^2+v_-^2)-\tfrac{17}{96}v_+v_-(v_+^2-v_-^2)+\mathcal{O}\left(v_{\pm}^5\right)
	\,.
\end{align}
Here $v=p/\ve,\ v'=p'/\ve',\ v_{\pm}=p_{\pm}/\ve_{\pm}$.
The leading terms in Eqs. \eqref{eq:asy_BS_low} and \eqref{eq:asy_PP_low} are in agreement with the results of Refs. \cite{sommerfeld1931beugung} and  \cite{Krachkov:2022fgt}, respectively, where the spectra of the non-relativistic  electron bremsstrahlung and of the pair production near threshold  were obtained exactly in the parameter $\frac{Z\alpha}{v}$.

\section{Graphs and checks}

Given the complexity of our results, it is important to have as many crosschecks as possible. First, we remind that the leading terms of the low- and high-energy asymptotics are in agreement with the results from the literature.

\begin{figure}
	\centering
	\includegraphics[width=0.7\linewidth]{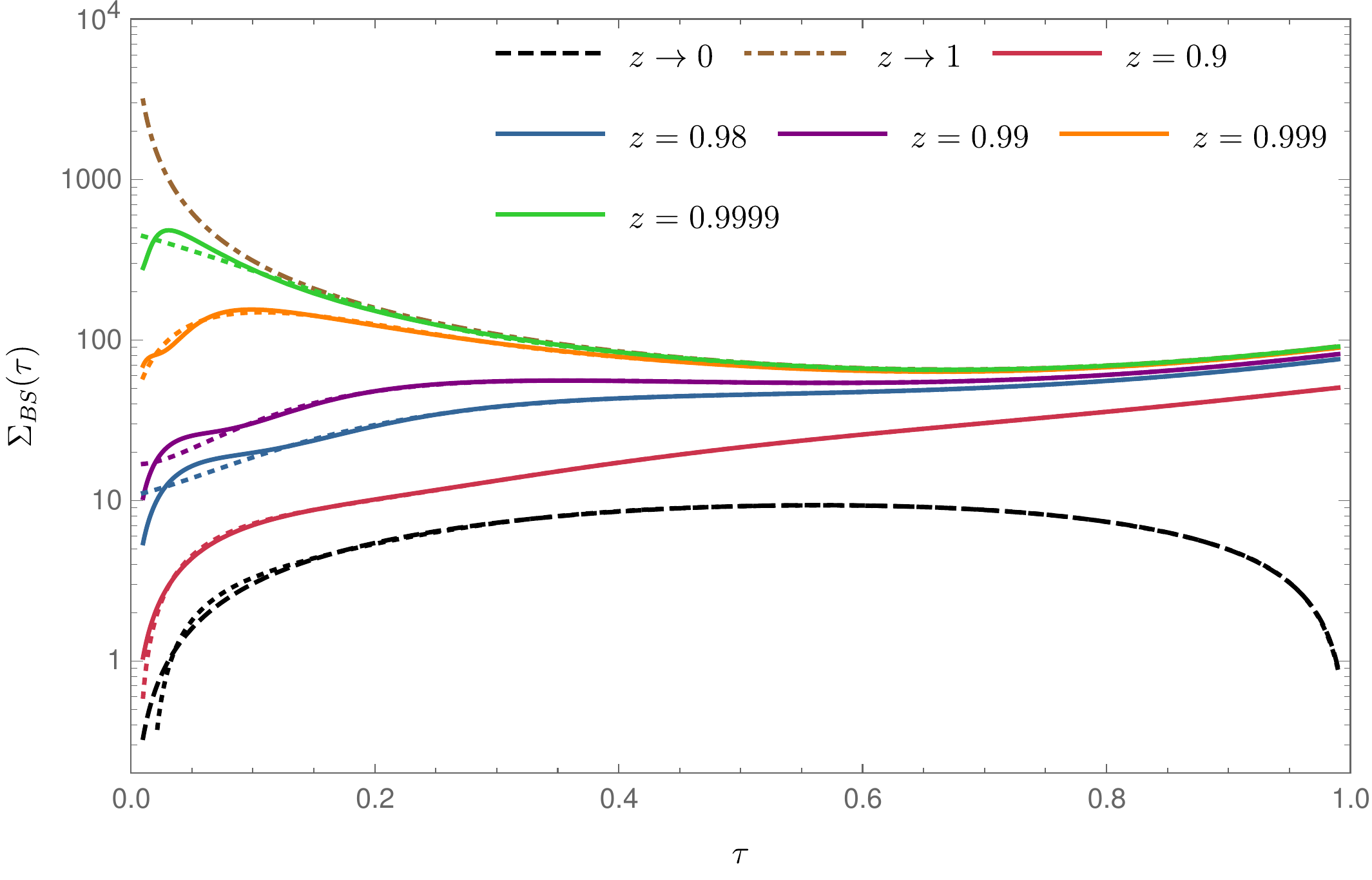}
	\caption{Function $\Sigma_{BS}(\tau)$ for several values of $Z$ and its comparison with low energy
		(dashed black curve) and high energy (dash-dotted brown curve) asymptotics. The solid line corresponds to our present result, the dotted line shows the fit of Ref. \cite{Lee2021}.}
	\label{fig:bsspectr}
\end{figure}

Another check for the bremsstrahlung process comes from the comparison with Ref. \cite{Lee2021}, where the energy loss in the bremsstrahlung was calculated in the order $\alpha (Z \alpha)^3$.
We have checked that the numerical integration of the bremsstrahlung spectrum \eqref{eq:BS} perfectly agrees with the result of Ref. \cite{Lee2021}.
In the same paper \cite{Lee2021} the authors approximately obtained the bremsstrahlung spectrum by fitting the moments
\begin{equation}
	K^{(n)} (Z,\varepsilon)=\int\limits_{0}^{\ve-m} \left(\frac{\ve'-m}{\ve-m} \right)^n \frac{\omega}{\varepsilon}\frac{d\sigma_C^{\BS}}{d\omega} d\omega\,.
\end{equation}
The comparison of our result with the results from  \cite{Lee2021} is presented in Fig. \ref{fig:bsspectr}, where the quantity
\begin{equation}
	\Sigma_{BS}(\tau)=\left(\frac{\alpha (Z\alpha)^3}{m^2}\right)^{-1}p \frac{\omega}{\ve}\frac{d\sigma_C^{\BS}}{d\tau}\,,
\end{equation}
is plotted as function of $\tau=\sqrt{\frac{\ve^\prime - m}{\ve - m}}$ for different $z$. A noticeable difference  at small $\tau$ is related to the uncertainty of the fitting functions basis chosen in Ref. \cite{Lee2021}.

For the pair production process the additional crosscheck comes from the comparison with the result of Ref. \cite{OMoOl1973}, where the exact spectrum and total cross section were obtained numerically for a few intermediate values of photon energy, $\omega-2m\sim m$, and atomic charge numbers.

\begin{figure}
	\centering
	\includegraphics[width=0.7\linewidth]{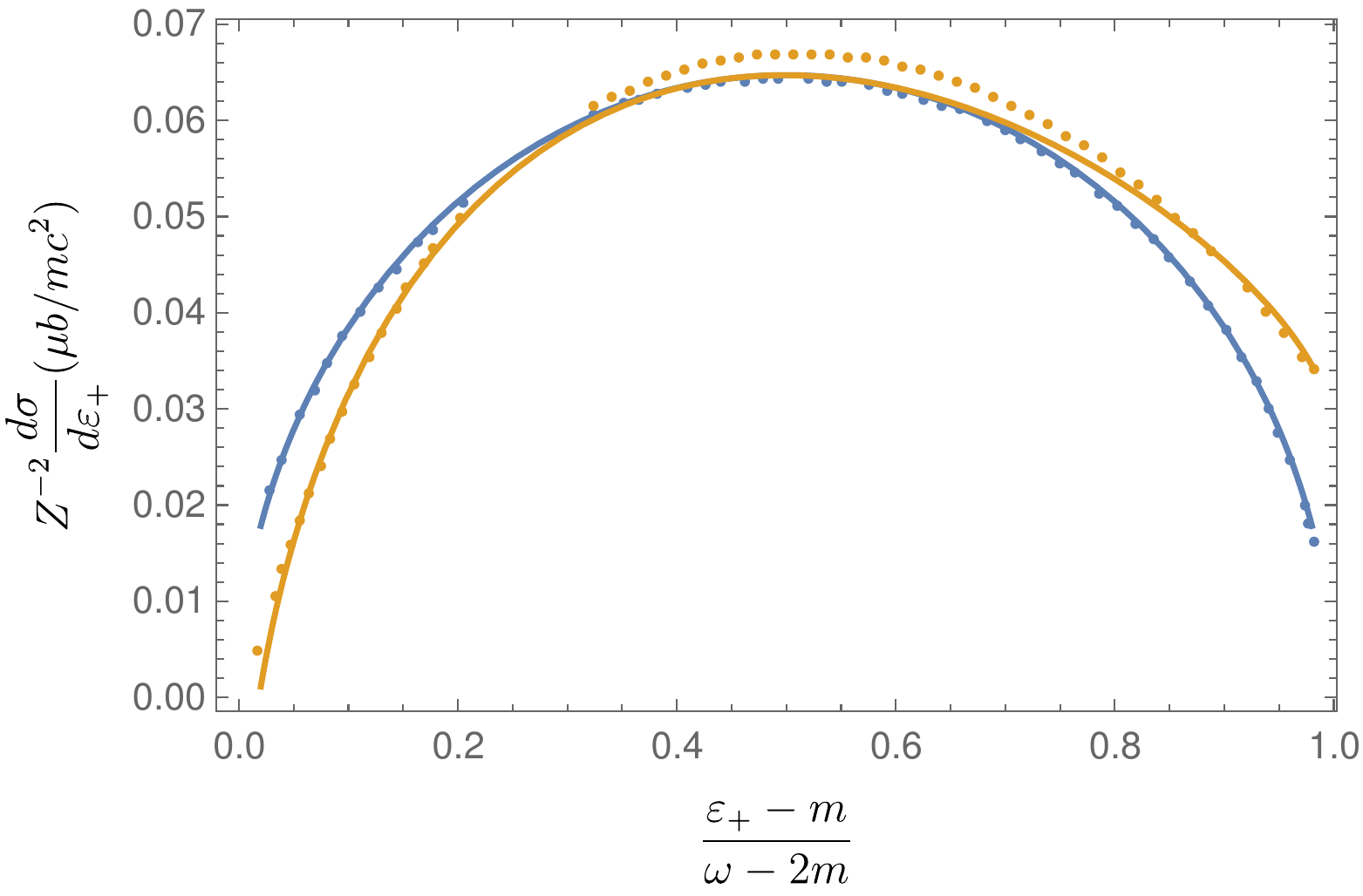}
	\caption{Positron energy spectra at $\om=3m$ for $Z=13$. Points correspond to exact numerical result from Ref.\cite{OMoOl1973}, solid line corresponds to the sum of Eqs. \eqref{eq:PP_born} and \eqref{eq:PP}. Blue points and curves corresponding to Born approximation are given for the reference.}
	\label{fig:ppspectr3}
\end{figure}

In Fig. \ref{fig:ppspectr3} the positron spectra at $\om=3m$ and $Z=13$ is presented. Points correspond to the numerical result of Ref.\cite{OMoOl1973}, line corresponds to the sum of Eqs.  \eqref{eq:PP_born} and \eqref{eq:PP}. We see that, apart from the central part, the solid curve already gives a good approximation. The deviation in the central part of the graph reaches $5\%$ and can be ascribed to higher orders in $Z\alpha$ not considered here.

\begin{figure}
	\centering
	\includegraphics[width=0.7\linewidth]{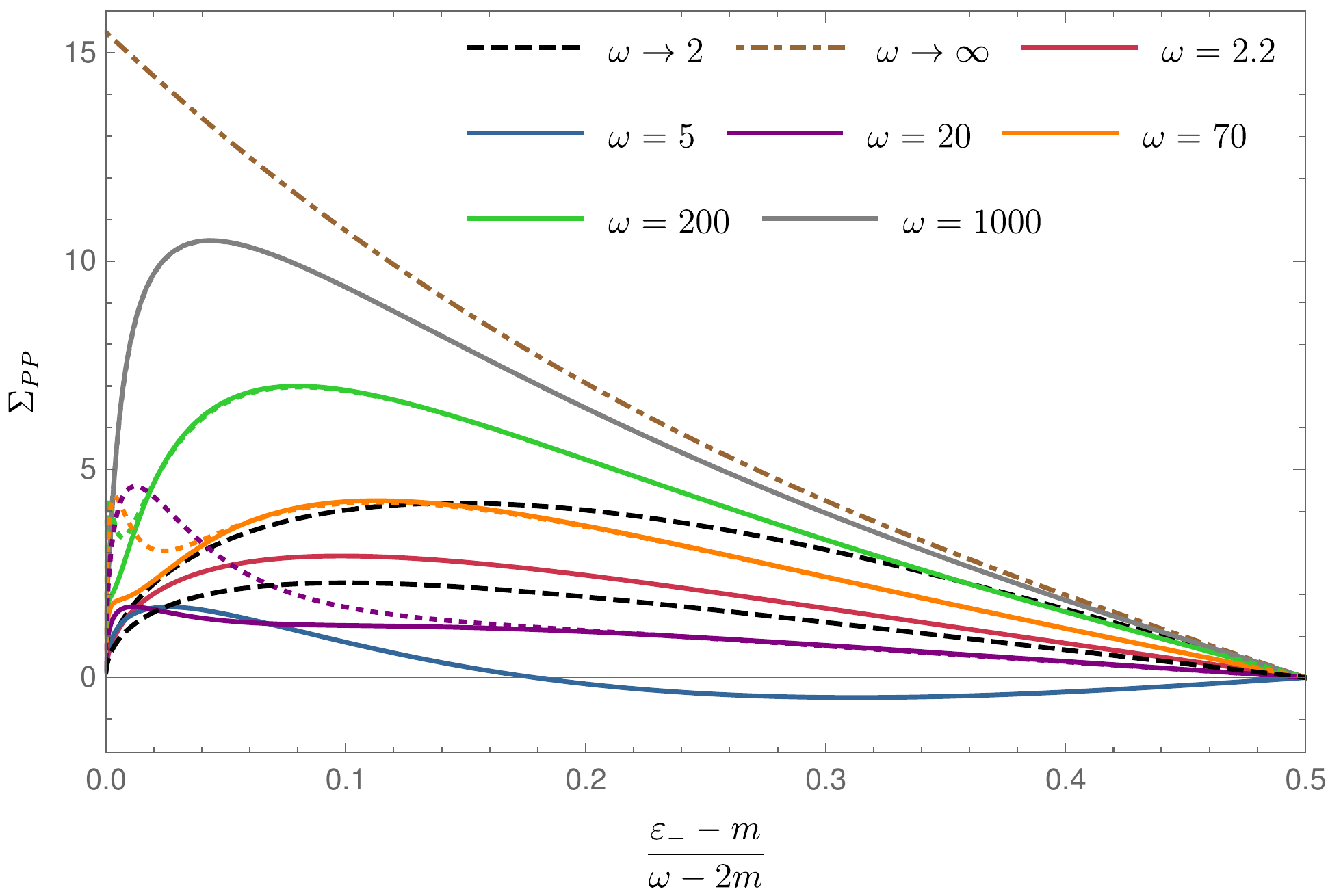}
	\caption{Function $\Sigma_{PP}$ for several values of $\omega$ and its comparison with leading low-energy (dashed black curve) and leading high-energy (dash-dotted brown curve) asymptotics. Color dashed lines correspond to the account of the first two terms in Eq. \eqref{eq:asy_PP}.}
	\label{fig:ppspectr}
\end{figure}

In Fig. \ref{fig:ppspectr} we plot the quantity
$$
\Sigma_{PP}=\frac{p_+p_-(p_++p_-)\omega^2}{\alpha(\Za)^3(\omega-2)^3}\frac{d\sigma_C^{\PP}}{d\ve_+}\,,
$$
as a function of $\frac{\varepsilon_--m}{\omega-2m}$ for different values of $\omega$. The convenience of this quantity is that it has finite low- and high-energy limits:

\begin{equation}
	\Sigma_{PP}^{\text{high}}=\frac{\pi^3}{4}(1-2\tilde{\tau})(2-3\tilde{\tau}(1-\tilde{\tau}))\,,\quad
	\Sigma_{PP}^{\text{low}}=\frac{16\pi}{3}(1-2\tilde{\tau})\sqrt{\tilde{\tau}(1-\tilde{\tau})}\,,
\end{equation}
where $\tilde{\tau}=\frac{\ve_--m}{\omega-2m}$.
As can be seen from Fig. \ref{fig:ppspectr}, the account of the $m^0$ term in Eq. \eqref{eq:asy_PP} essentially improves the precision of the high-energy approximation.

\section{Conclusion}
In the present paper we have calculated the first Coulomb corrections to the spectra of the electron bremsstrahlung and electron-positron pair photoproduction in the Coulomb field. These corrections give the leading contribution to the charge asymmetry in these processes. Our results are expressed in terms of dilogarithms and simpler functions. We observe that, in contrast to the Born spectra and the leading high-energy asymptotics, the obtained results for these two processes are not related by the crossing symmetry rules. We provide an explicit example of a pair of master integrals which can not be related by an analytic continuation from bremsstrahlung channel to pair production channel. Comparison with the results available in the literature, such as high- and low-energy asymptotics, asymmetry in the total energy loss and numerical results for pair production spectra, shows perfect agreement. We provide \textit {Mathematica} files containing the obtained formulas in computer-readable form.
\acknowledgments We are grateful to V.S. Fadin and A.I. Mistein for the interest to the work and fruitful discussions. The work has been supported by Russian Science Foundation under grant 20-12-00205.

\appendix

\section{Coefficients $c_k$ and $\tilde{c}_k$}
The coefficients $c_{1-12}$ in Eq. \eqref{eq:BS} have the form
\begin{multline}
	c_1=
	\frac{2 \ve }{p}
	\,,\quad
	c_2=
	\frac{2 \left(p'^4-p^4\right)}{p^2 p'^2}-\frac{4 m^2 \ve' \om'}{p'^4}
	\,,\quad
	c_3=
	\frac{4 \ve'^3 \om'+2 p'^2 \om'^2}{p'^4}
	\,,\\
	c_4=
	\frac{\ve+\ve'}{m} \left(\frac{2 \om'}{\ve'+m}-\frac{2 \om'}{\ve +m}+3\right)
	,\quad
	c_5=
	\frac{2\left(p^2-p'^2\right)^2}{p^2 p'^2}
	\,,\quad
	c_6=
	\frac{2 \om' \left(2 \ve  p'-p \ve'\right) \left(\ve  \ve'+p'^2\right)}{p p'^4}
	\,,\\
	c_7=
	\frac{2 \ve' \om' \left(\ve  \ve'+p^2\right)}{p^3 p'}
	\,,\quad
	c_8=
	\frac{2 \ve'}{p'}
	\,,\quad
	c_9=
	-\frac{8 \ve  \ve'^2\, \left(\ve  \ve'-p p'\right)}{3 m^2 p p'^2}
	-\frac{2 \ve' \left(p \ve'-\ve  p'\right)^2 \left(\ve ^2+p'^2\right)}{m^2 p^2 p'^3}\\
	+\frac{2 \left(3 \ve'^2-5 \ve ^2\right) \ve'}{3 p p'^2}+\frac{2 \ve  \ve'^2\,\left(p^2+p'^2\right)}{p^2 p'^3}
	\,,\quad
	c_{10}=
	\frac{2 \ve'^2}{p^2}
	\,,\quad
	c_{11}=
	\frac{2 \left(3 \ve -\ve'\right) \ve'^2 \om'}{p'^4}+\frac{\ve'^2+\ve \ve'}{p'^2}\\-\frac{2 \ve \ve' p }{p'^3}
	\,,\quad
	c_{12}=
	\frac{\om' \left(6 \ve  \ve'-3 \ve'^2+m^2\right)}{p'^3}+\frac{2 \left(p \ve'-\ve  p'\right) \left(3 \ve ^2+3 p'^2+2 p p'\right)}{3 m^2 p p'}-\frac{2 \ve  p}{p'^2}
	\,.
\end{multline}

The coefficients $\tilde{c}_{1-9}$ in Eq. \eqref{eq:PP} have the form

\begin{multline}
	\tilde{c}_1=
	\frac{2\ve_-}{p_-}
	\,,\quad
	\tilde{c}_2=
	\frac{2 \om\,  \left(\left(\ve_--\ve_+\right) \ve_-^2+m^2 \om \right)}{p_-^4}
	\,,\quad
	\tilde{c}_3=
	\frac{2 \left(p_+^2-p_-^2\right) \left(p_-^2+p_+^2-m^2\right)}{p_-^2 p_+^2}
	\\
	-\frac{2 m^2\, \left(p_+^4-p_-^4\right) \left(\ve_- \ve_++m^2\right)}{p_-^4 p_+^4}
	\,,\quad
	\tilde{c}_4=
	\frac{\left(\ve_--\ve_+\right)}{2 m} \left(3-  \frac{2 \om}{\ve_++m}-\frac{2 \om}{\ve_-+m}\right)
	\,,\quad\\
	\tilde{c}_5=
	\frac{2 \om\,  \left(\ve_+ p_--\ve_- p_+\right) \left(p_+^2-\ve_- \ve_+\right)}{p_- p_+^4}
	\,,\quad
	\tilde{c}_6=
	\frac{2\left(\ve_- \ve_++m^2\right) \left(m^2+p_-^2+p_+^2\right)}{p_-^3 p_+}
	\,,\quad\\
	\tilde{c}_7=
	\frac{\ve_+ p_--\ve_- p_+}{p_+p_-} \left(\frac{4 \left(3 \ve_-^2+\ve_+ \ve_-\right)}{3 m^2}+\frac{2\om\, \ve_+}{p_-^2}\right)+\frac{7 \ve_- \ve_++5 m^2}{p_-^2}
	\\
	+\frac{2\left(\ve_- \ve_++m^2\right) \left(3 \ve_- \ve_++m^2\right)}{p_-^4}
	\,,\quad
	\tilde{c}_8=
	\frac{2 \left(\ve_+ p_- \left(\ve_--p_+\right)-\ve_-^2 p_+\right)}{p_+^3}
	\,,\quad\\
	\tilde{c}_9=
	\frac{2 \ve_- \ve_+\left(\ve_+ p_--\ve_- p_+\right)}{m^2 p_-^2}
	-\frac{4\ve_- p_+}{3 m^2}
	+\frac{2p_+\left(3 \ve_- \ve_++m^2\right)}{p_-^3}
	+\frac{5p_+}{ p_-}
	\,.
\end{multline}
\label{sec:appendix}
\bibliographystyle{JHEP}
\bibliography{BSPPrefs}
\end{document}